# Fe and Mg Isotope compositions Indicate a Hybrid Mantle Source for Young Chang'E 5 Mare Basalts

Yun Jiang[1,2,†], Jinting Kang[3,4,2,†], Shiyong Liao[1,2], Stephen M. Elardo[5], Keqing Zong[6], Sijie Wang[3], Chang Nie[1], Peiyi Li[1], Zongjun Yin[7], Fang Huang[3,4,2], Weibiao Hsu[1,2,4]

**Abstract**

The Chang'E 5 (CE-5) samples represent the youngest mare basalt ever known and provide an access into the late lunar evolution. Recent studies have revealed that CE-5 basalts are the most evolved lunar basalt, yet controversy remains over the nature of their mantle sources. Here we combine Fe and Mg isotope analyses with a comprehensive study of petrology and mineralogy on two CE-5 basalt clasts. These two clasts have a very low Mg# (~29) and show similar Mg isotope compositions with Apollo low-Ti mare basalts as well as intermediate $TiO_2$ and Fe isotope compositions between low-Ti and high-Ti mare basalts. Fractional crystallization or evaporation during impact cannot produce such geochemical signatures which otherwise indicate a hybrid mantle source that incorporates both early- and late-stage lunar magma ocean (LMO) cumulates. Such a hybrid mantle source would be also compatible with the KREEP-like REE pattern of CE-5 basalts. Overall, our new Fe-Mg isotope data highlight the role of late LMO cumulate for the generation of young lunar volcanism.

1.  **Introduction**

The Procellarum KREEP (Potassium, Rare Earth Elements, and Phosphorous) Terrane (PKT) accounts for only ~ 16% of the lunar surface area, but more than 60% (by area) of mare basalts are located within this terrane, with volcanic activities extending from as early as 4.2 Ga (Taylor et al. 1983) to as late as ~ 2.0 Ga (Che et al. 2021; Li et al. 2021). Crater-counting chronology even suggests that the youngest volcanism in PKT could approach 1.0 Ga (Hiesinger et al. 2000), indicating that the duration of basalt eruptions is as long as 3 billion years. The driving force for the prolonged lunar volcanism has long been a pending issue. The PKT exhibits the enrichments of heat-producing elements (e. g., 3 to 12 μg/g Th) (Lawrence et al. 1998), leading to a possible mechanism that heat produced through long-lived radioactive decay may have sustained magmatic activity on the nearside of the Moon (Borg 2004). Other mechanisms could be also possible, such as tidal heating (Harada et al. 2014) and an insulating megaregolith layer (Ziethe, Seiferlin, & Hiesinger 2009). Chang'E-5 (CE-5) basalts were sampled from the PKT, marking the youngest

† These authors contributed equally.
[1] CAS Key Laboratory of Planetary Sciences, Purple Mountain Observatory, Chinese Academy of Sciences, Nanjing 210023, China. Corresponding author wbxu@pmo.ac.cn
[2] Center for Excellence in Comparative Planetology, Chinese Academy of Sciences, Hefei 230026, China
[3] CAS Key Laboratory of Crust-Mantle Materials and Environments, School of Earth and Space Sciences, University of Science and Technology of China, Hefei 230026, China. Corresponding author kjt@ustc.edu.cn
[4] Deep Space Exploration Laboratory/School of Earth and Space Sciences, University of Science and Technology of China, Hefei 230026, China
[5] The Florida Planets Lab, Department of Geological Sciences, University of Florida. Gainesville, FL 32506, USA.
[6] State Key Laboratory of Geological Processes and Mineral Resources, School of Earth Sciences, China University of Geosciences, Wuhan 430074, China
[7] State Key Laboratory of Palaeobiology and Stratigraphy, Nanjing Institute of Geology and Paleontology & Center for Excellence in Life and Paleoenvironment, Chinese Academy of Sciences, Nanjing 210008, China

(~ 2.0 Ga) known lunar samples to date (Che et al. 2021; Li et al. 2021). Their petrology and geochemistry can shed new light on the driving mechanism of the production and eruption of young lava flows.

Identification of the source lithology of mare basalts plays a fundamental role in the understanding of lunar magmatic process and the origin of young lunar volcanism. However, mare basalts have generally undergone various degrees of fractional crystallization before eruption (Neal 1992), particularly for CE-5 basalts, which are the most evolved mare basalt to date (Tian et al. 2021). The complicated fractional crystallization history hinders the application of major and trace elements to retrieve their source characteristics. Tian et al. (2021) first pointed out that although CE-5 basalts exhibit a KREEP-like REE pattern, the depleted Sr-Nd isotope compositions indicate a non-KREEP origin, probably with olivine-dominated cumulate as the source materials. In contrast, Zong et al. (2022) argued that if the mantle source of CE-5 basalts is clinopyroxene-rich, a small fraction (1 to 1.5%) of KREEP components would be reconciled with the depleted Sr-Nd isotope signatures. Recently, Su et al. (2022) suggested that the clinopyroxene-rich cumulates crystallized in the late-stage of lunar magma ocean (LMO) solidification is more fusible than the early formed olivine-dominated mantle. The involvement of the late stage cumulates by overturn can lower the mantle melting point and enable the melting of lunar interior at 2.0 Ga. Thus, identification of the lithology of CE-5 mantle source is the key issue.

Iron and magnesium isotope systems have been investigated for lunar basalts and show an isotope dichotomy in low- and high-Ti basalts (Liu et al. 2010; Poitrasson et al. 2004; Sedaghatpour et al. 2013; Sossi & Moynier 2017; Weyer et al. 2005; Wiechert & Halliday 2007). Both experimental and theoretical studies indicate that Fe-Mg isotope composition of lunar basalts are less affected by olivine fractional crystallization (Huang et al. 2013; Prissel et al. 2018). Instead, as pyroxene shows lighter Fe and heavier Mg isotope compositions than olivine, Sedaghatpour & Jacobsen (2019) suggest the Fe-Mg isotope dichotomy in mare basalts should reflect the lithology heterogeneity of their mantle source, with a clinopyroxene-rich, late-stage cumulate for high-Ti basalts and an olivine- or orthopyroxene-rich, early-stage cumulate for low-Ti basalts. Therefore, Fe-Mg isotope systems could be applied as tools in deciphering the source lithology of mare basalts. In addition, previous studies mainly focused on single isotope system of lunar basalts, either Fe or Mg isotope. In this work, we combine Fe-Mg isotope investigations with a comprehensive study of petrology, mineralogy, bulk major, trace elemental, and radiogenic Sr isotope compositions of CE-5 basalt clasts, to conclude that young CE-5 mare basalts possess a hybrid mantle cumulate source, which may be pivotal in prolonging volcanic activity on the Moon.

2. **Samples and Analysis Procedure**

In this work, two CE-5 basalt clasts (CE-5-01 and CE-5-02) are from the batch sample (CE5C0800YJYX038) allocated by the China National Space Administration. We first carried three-dimensional tomography observations to make sure that they are lava flow fragments, rather than impact melts. Representative slivers were then handpicked from each clast and prepared for petrography and mineral chemistry. Remaining basaltic fragments were dissolved for bulk-rock major, trace elements, Fe-Mg-Sr isotope analyses. Detailed analytical methods and data are given in the Appendix.

## 3. Results and Discussion

### 3.1. *The Unique CE-5 Mare Basalts*

Both 3D tomography and 2D petrologic observations show that the two clasts have a sub-ophitic texture (Figure A1 in Appendix) and a remarkably similar pyroxene, olivine, and plagioclase mineral chemistry to those reported by Tian et al. (2021) and Jiang et al. (2022) (Figure A2). They also have the same low initial $^{87}Sr/^{86}Sr$ value (0.69977, Appendix). This suggests that they are representative of the local mare basalt at the CE-5 landing site (designated as unit Em4/P58). It should be noted that pyroxene compositions display a fractional crystallization trend distinct from that in the low-Ti basalt NWA 10597, but intermediate between those in low-Ti and high-Ti basalts (Robinson, Treiman, & Joy 2012) (Figure A2).

Using the $TiO_2$-calibrated $K_d$ of 0.32 (Delano 1980), the calculated Fo content of olivine that is in equilibrium with a whole rock Mg# (atomic Mg/ (Mg + $Fe^{2+}$) (29.6 and 28.2, Table A1) is approximately 56.7 and 55.2, which is close to the highest measured Fo content in CE-5-01 and CE-5-02 ($Fo_{52.7}$ and $Fo_{58.3}$), indicating no significant crystal accumulation. We thus conclude that the measured bulk composition of clasts is representative of a bulk liquid that crystallized as a relatively closed system. According to the bulk chemistry classification scheme of Apollo 15 mare basalts developed by Rhodes & Hubbard (1973), CE-5 basalts should be an olivine normative basalt ($SiO_2$ < 46 wt.%, FeO > 21 wt.%, and $TiO_2$ > [3.62 – 0.157*MgO]). In terms of major elements, two CE-5 basalts exhibit a very low Mg# of 29 (average value), which is indicative of a more evolved basalt than all known Apollo mare basalts (Figure A3). The highly evolved property is consistent with the observed high abundance of late stage mesostasis (Figure A1). They have $TiO_2$ contents (average 5.75 wt.%) lower than Apollo 11 (hereafter referred to as A11) and A17 high-Ti basalts (8.0–13.3 wt.%), but slightly higher than A12 and A15 low-Ti basalts (1.6–4.5 wt.%) (Figure A3). Previous studies have shown that low-Ti and high-Ti mare basalts cannot be related through crystal-liquid fractionation and cannot have been derived from the same mantle source through variable degrees of partial melting (Papike 1976). Thus, the intermediate $TiO_2$ contents of CE-5 basalts likely reflects a unique mantle source. In addition, the MgO content of CE-5 in this work (average 5.63 wt.%) is consistent with those of Tian et al. (2021) and Zong et al. (2022) but higher than that (4.4 ± 0.8 wt.%) of Su et al. (2022) (Figure 1). Our $Al_2O_3$ content (average 9.68 wt.%) is lower than that (13.2 ± 1.5 wt.%) of Su et al. (2022) which is estimated based on the petrographic thick sections. Su et al. (2022) suggested that CE-5 basalts first underwent olivine crystallization and then clinopyroxene crystallization. Here, we performed a reverse fractional crystallization modeling, using the primary magma compositions in Su et al. (2022). The result shows that olivine fractional crystallization solely can explain the $Al_2O_3$-MgO content of CE-5 basalts in this work (Figure 1). Thus, unlike the result of Su et al. (2022), the two clasts in this work represent a less evolved CE-5 basalt.

CE-5 basalts are the most incompatible trace elements-enriched mare basalt yet found, with elevated REE [e. g., (La) $_N$ = 158], Th (5.1 μg/g) and a pronounced negative Eu-anomaly (Eu/Eu* = 0.5(Sm + Gd)] $_N$ of 0.49 (Figures A3 and A4). They have an extremely high (La/Yb) $_N$ compared to the range observed in other mare basalts (≥ 2.8 vs. 0.3–2.0) (Figure A3) and display a KREEP-like REE pattern (Figure A4). Tian et al. (2021) initially suggested that the REE pattern of CE-5 basalts can be explained by 43–78% fractional crystallization, with a hypothesized crystallization assemblages of 5–10% olivine, 25–59% augite, 2–3% pigeonite, and 6–11% plagioclase. Differently, our samples mainly underwent olivine crystallization as suggested by the reverse fractional crystallization modeling (Figure 1). Since Eu is highly incompatible in olivine (McKay 1986), the crystallization of olivine cannot produce such a pronounced negative Eu-anomaly. Therefore, the KREEP-like REE pattern should reflect other process such as a source incorporating cumulates that were formed in LMO after extensive fractional crystallization of plagioclase.

### 3.2. A Hybrid Mantle Source For CE-5 Basalts

Our two CE-5 basalt clasts show excellent reproducibility in Fe and Mg isotopic compositions (Figure 2). The average $\delta^{57}$Fe (0.161 ± 0.010‰) is at the upper end of low-Ti basalts (0.125 ± 0.051‰) and is lower than that of high-Ti basalts (0.264 ± 0.093‰) (Liu et al. 2010; Poitrasson et al. 2004; Sossi & Moynier 2017; Weyer et al. 2005). The average $\delta^{26}$Mg (-0.264 ± 0.015‰) is in the range of low-Ti lunar basalts (-0.263 ± 0.115‰) and is heavier than that of high-Ti basalts (-0.474 ± 0.198‰) (Sedaghatpour & Jacobsen 2019; Sedaghatpour et al. 2013).

Previous studies on lunar samples have identified a bimodal distribution of Fe-Mg isotope compositions for low- and high-Ti basalts (e.g., Liu et al. 2010; Poitrasson et al. 2004; Sedaghatpour et al. 2013; Sossi & Moynier 2017; Weyer et al. 2005; Wiechert & Halliday 2007) (Figure 2). Such variations may reflect several factors, including: 1) evaporative loss of light isotopes during impacts or accretion (Hin et al. 2017; Poitrasson et al. 2004); 2) kinetic isotope fractionation induced by chemical diffusion (Prissel et al. 2018; Sossi & Moynier 2017); 3) magma evolution (Chen et al. 2021); 4) source heterogeneity that resulted from LMO differentiation (Sedaghatpour & Jacobsen 2019; Sossi & Moynier 2017). The CE-5 basalts show a slightly heavier Fe isotope composition (0.161 ± 0.010‰), but a similar Mg isotope composition (-0.264 ± 0.015‰) than low-Ti basalt ($\delta^{57}$Fe: 0.125 ± 0.051‰ and $\delta^{26}$Mg: -0.263 ± 0.115‰) (Figure 2). It is unlikely that the heavy Fe isotope composition of CE-5 basalts originated from a volatilization event, because the two studied clasts do not display significant impact features based on 3D and 2D petrologic observations. Wang et al. (2015) found extremely low $\delta^{57}$Fe (-0.53‰) in lunar dunite and invoked diffusion driven kinetic fractionation to explain such a signature because light isotopes diffuse faster than heavy isotopes. Disequilibrium fractionation between olivine and melt may explain the heavy Fe isotope composition in CE-5 basalt. However, Fe and Mg diffusion are coupled in olivine and kinetic isotope effect will produce an opposite Fe-Mg isotope fractionation trend (Kin I Sio & Dauphas 2017; Wang et al. 2015). This is inconsistent with the unfractionated Mg isotope composition observed for CE-5 basalts, and hence ruling out the possibility of kinetic fractionation. Moreover, studies on terrestrial basalts have shown that fractional crystallization can elevate the $\delta^{57}$Fe of residual melts (e.g., Chen et al. 2021). However, experiments conducted on basaltic magma similar to lunar basalt compositions and oxygen fugacity have indicated negligible Fe isotope fractionation during olivine crystallization (Prissel et al. 2018). As samples in this study mainly underwent olivine fractional crystallization, the heavy Fe isotope compositions of the CE-5 basalts should not result from magma evolution.

Alternatively, the Fe-Mg isotope signature of CE-5 basalts may reflect a different mantle source from Apollo low-Ti basalts. The mantle of lunar basalts are cumulates formed in distinct stages of LMO crystallization. Olivine crystalized in the early stage of LMO crystallization and shows heavier Fe and lighter Mg isotope compositions than pyroxene that crystallized in the late stage of LMO crystallization (Huang et al. 2013; Nie et al. 2021). Therefore, Sedaghatpour & Jacobsen (2019) attributed the $\delta^{26}$Mg-$\delta^{57}$Fe bimodal distribution in lunar basalts as remelting of distinct LMO cumulate sources. Here, we modeled Fe-Mg isotope fractionation during LMO differentiation using the model of Charlier et al. (2018). As suggested by Snyder, Taylor, & Neal (1992), LMO solidification likely starts as equilibrium crystallization and changes to fractional crystallization at some degree of crystallization. In our model, the transition of equilibrium to fractional crystallization is assumed to occur at 40 Per Cent Solid (PCS), in agreement with previous studies that solidification of the LMO was dominated by fractional crystallization (Rapp & Draper 2018; Suckale, Elkins-Tanton, & Sethian 2012). More details regarding the parameters used in this model are presented in Appendix A.4.

Our predicted MgO and FeO contents for LMO cumulates and Fe-Mg isotope compositions of cumulate and residual melt are shown in Figure 3. The Fe and Mg isotope compositions are unfractionated during the olivine crystallization stage (< 53 PCS) because petrology experiments and theoretical calculations show that olivine has similar Fe and Mg isotope compositions with basaltic melts (e. g., Nie et al. 2021; Wang et al. 2023). The modeling results further suggest that the pyroxene crystallization controls the Fe-Mg isotope evolution of LMO. The predicted Fe-Mg isotope trend in this work is slightly different from the results in Sedaghatpour & Jacobsen (2019) because we used the most recent constrained equilibrium isotope fractionation factors and considered the cooling effect on equilibrium fractionation factors. In addition, we applied a more recent LMO crystallization sequence of Charlier et al. (2018), while Sedaghatpour & Jacobsen (2019) used the crystallization sequence of Snyder, Taylor, & Neal (1992). Nevertheless, we come to the same point as Sedaghatpour & Jacobsen (2019) did that early LMO cumulates have lower FeO and $\delta^{57}$Fe, higher MgO and $\delta^{26}$Mg than late LMO cumulates.

A batch melting model is applied to calculate the Fe-Mg isotope compositions of partial melts derived from 50, 70, 86 and 98 PCS cumulates at 1–20% melting degree (Figure 4). Zong et al. (2022) indicated that CE-5 basalts may be formed by 3% partial melting of 86 PCS cumulate. The predicted melts derived from 86 PCS cumulate have $\delta^{57}$Fe of -0.05 to 0.04‰ and $\delta^{26}$Mg of -0.30 to -0.31‰, inconsistent with the observed $\delta^{57}$Fe (0.161 ± 0.010‰) and $\delta^{26}$Mg (-0.264 ± 0.015‰) of CE-5 basalts in this work. Moreover, 86 PCS cumulate should be located at a relatively shallow depth (100–150 km) if assuming a 600 km depth LMO (Charlier et al. 2018). Reheating of a shallow mantle to the solidus temperature (> 1100°C) at ~ 2.0 Ga conflicts with the strong and cool upper mantle required to support lunar mascons (Hess & Parmentier 2001). Therefore, the CE-5 basalts may not be produced by partial melting of a cumulate layer formed at 86 PCS.

The detailed major elemental investigations by Su et al. (2022) show that the primary magma of CE-5 basalts has more CaO and $TiO_2$ than that of low-Ti basalts. This may reflect an olivine-dominated cumulate source incorporated pyroxene-rich cumulate that was lately formed in LMO. As shown by the LMO modeling (Figure 3), late cumulates have significantly lower MgO content and higher FeO content than early cumulates. Thus, the mixture between early and late cumulates may explain the heavy Fe isotope and unfractionated Mg isotope signature of CE-5 basalts. Here, we conducted two-endmember mixing modeling to evaluate the possible source region of CE-5 basalts. As the primary magma of CE-5 basalt show higher $TiO_2$ than low-Ti basalts (Su et al. 2022), the late cumulate should contain abundant ilmenite. In the LMO model of Charlier et al. (2018), ilmenite is saturated after 97 PCS and hence we choose 98 PCS cumulate as the late cumulate endmember. We tested the scenarios of 98 PCS partial melts (1–20% degree) mixed with partial melts (1–20%) of 50, 70, and 86 PCS. The modeling results show that 98 PCS melts mixed with 50 PCS melts can well explain the Fe-Mg isotope compositions of CE-5 basalts and the contribution of 98 PCS melts is 20–30%. The 50 PCS cumulate is composed by 100% olivine and the 98 PCS cumulate is clinopyroxene-rich with 44% pigeonite + 33% augite +23% ilmenite if assuming all plagioclase floated to the surface of LMO. Therefore, our new Fe-Mg isotope data indicate that a late-stage formed, clinopyroxene-rich cumulate should exist in the source region of CE-5 basalts.

An important but unsettled question is how the KREEP-like REE pattern of CE-5 basalt is formed, since its mantle source contains few KREEP-like materials (< 1%) (Tian et al. 2021; Zong et al. 2022). To test whether a hybrid cumulate source can explain the KREEP-like REE pattern, we calculated the REE contents of partial melts derived from hybrid LMO cumulates. LMO cumulates contain not only mafic minerals but also a small proportion of entrained plagioclase (~7%) and trapped instantaneous residual liquid (TIRL, ≤ 5%) (Van Orman & Grove 2000). These two components will not significantly affect the Fe-Mg isotope variations but can significantly affect incompatible trace element distribution. In addition, the REE pattern of melt is sensitive to the melting degree as well as the mixing ratio of the cumulate source endmember. Although those uncertainties increased the complexity of the calculations, our results show that the KREEP-like REE pattern can be reproduced by selecting reasonable parameter combinations (Figure 5). For instance, partial melts of 50 PCS +3% TIRL source mixed with melts of 98 PCS + 3% TIRL source at a melting degree of 0.5% and in a mixing ratio of 70: 30 (mass ratio, Figure 5a); melts of 50 PCS + 5% TIRL mixed with melts of 98 PCS + 5% TIRL at a melting degree of 1% and in a ratio of 70: 30 (Figure 5b). Therefore, the formation of KREEP-like REE pattern of CE-5 basalt is not necessary to invoke the incorporation of KREEP materials in the source region and can be explained by low degree melting of a hybrid cumulate source.

New data in this work support that the CE-5 basalts may represent a mixture of melts from early formed and late formed cumulates. This may occur through two possible mechanisms: 1) partial melts of early formed cumulate assimilated late formed cumulate during melt ascent; or 2) late formed cumulate sank into the deep mantle, partially melted, and mixed with early formed cumulate. The first possibility requires the assimilation of lunar's upper mantle or lower crust, which seems difficult at ~ 2.0 Ga because the lithosphere should have been cold and stagnant (Hess & Parmentier 1995; Laneuville, Taylor, & Wieczorek 2018; Shearer et al. 2006). The second possibility is consistent with sinking of a dense ilmenite-rich cumulate during mantle overturn (Hess & Parmentier 1995). The downwelling late cumulates may have produced discrete regions of mantle enriched in ilmenite and clinopyroxene. These ilmenite-clinopyroxene cumulates are less refractory, with solidus temperature a few hundreds of degrees lower than olivine-orthopyroxene cumulates (Wyatt 1977). These cumulates will preferentially melt and ascend into surrounding refractory cumulate to form a hybrid mantle (Elkins-Tanton et al. 2002; Li et al. 2019). As shown by the phase equilibrium modeling of Su et al. (2022), such a hybrid mantle is more fusible than olivine-orthopyroxene dominated mantle. Therefore, the late cumulate migrate downward by mantle overturn may lower the mantle melting point and enables the prolonged lunar volcanism until ~ 2.0 Ga.

4. **Conclusion**

CE-5 basalt clasts in this work represent the local mare basalt at the CE-5 landing site. They are an ilmenite-bearing, clinopyroxene-rich olivine-normative mare basalt, characterized by low Mg# (average 29), high FeO (24.7 wt.%) and Th (5.1 μg/g), intermediate TiO$_2$ concentrations (5.75 wt.%) and δ$^{57}$Fe (0.161 ± 0.010‰) between Apollo low-Ti and high-Ti mare basalts. They have a similar Mg isotope composition (-0.264 ± 0.015‰) with low-Ti basalts.

CE-5 basalts are the most evolved basalt to date on the Moon, exhibiting the most incompatible trace elements-enriched [e. g., (La)$_N$ = 158] and KREEP-like REE pattern with a pronounced negative Eu anomaly (0.49). Our new Fe-Mg isotope data indicate that the young CE-5 mare basalts possess a hybrid mantle cumulate source that incorporates both early- and late-stage LMO cumulates, which may play an important role in the generation of the late lunar volcanism.

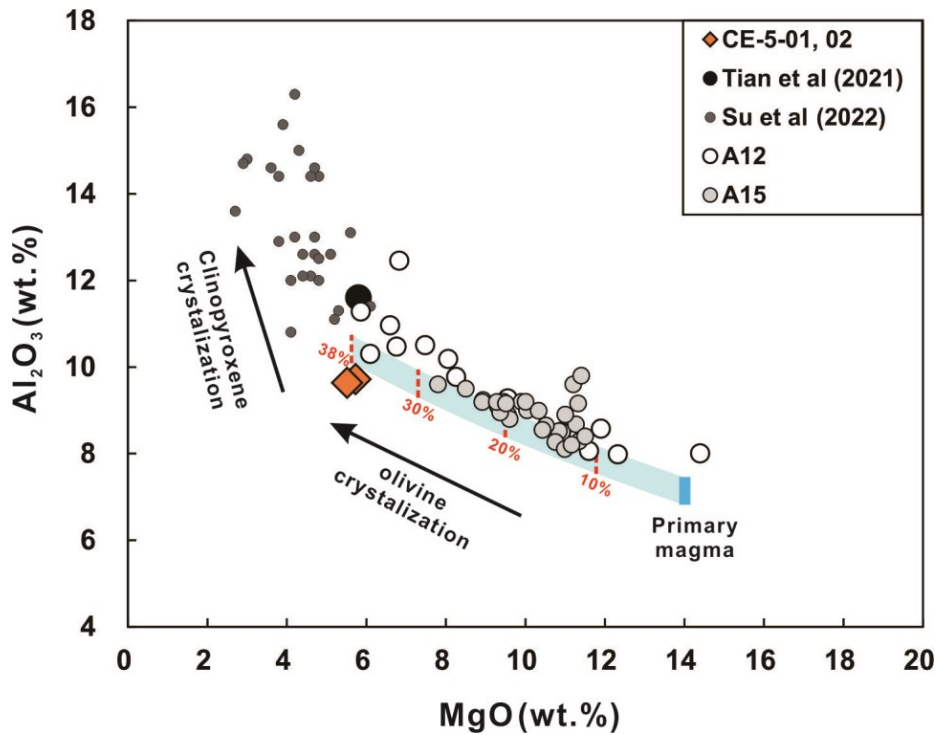

**Figure 1. Major element variation diagrams Al$_2$O$_3$ vs. MgO in wt.% for CE-5 basalts and the ideal fractional crystallization path of CE-5 parental liquids calculated by 1% steps of olivine addition.** The numbers aside the curve denote the adding degree. CE-5 basalt (Su et al. 2022; Tian et al. 2021), A12 and A15 basalts (Clive Neal's Mare Basalt database, https://www3.nd.edu/~cneal/Lunar-L/) are plotted. The CE-5 basalt may undergo 38% fractional crystallization of olivine if assuming its primary magma have similar MgO and Al$_2$O$_3$ with the estimated value in Su et al. (2022). The olivine composition for each increment in equilibrium with the liquid was recalculated following the equations presented in Charlier et al. (2018).

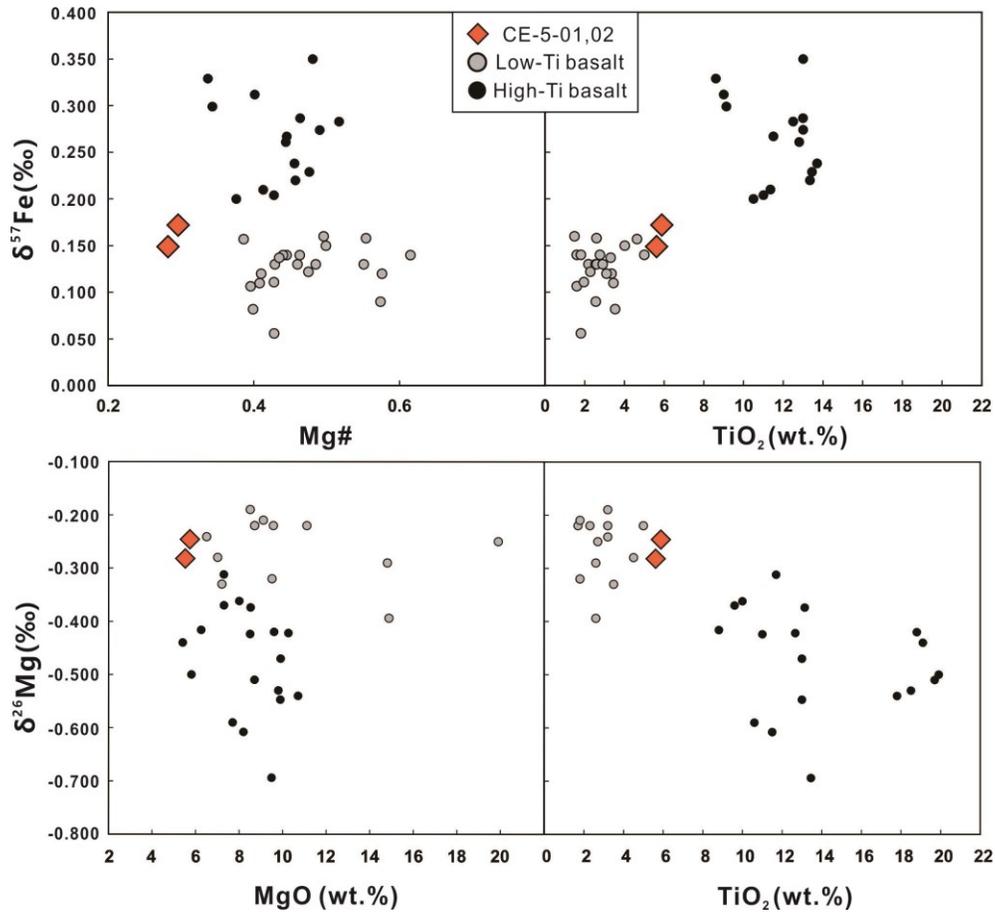

**Figure 2. Fe and Mg isotopes of CE-5-01 and CE-5-02.** The $\delta^{57}$Fe of CE-5 basalts are at the upper end of low-Ti basalts (0.125 ± 0.051‰) and are lower than high-Ti basalts (0.264 ± 0.093‰) (Liu et al. 2010; Poitrasson et al. 2004; Sossi & Moynier 2017; Weyer et al. 2005). The $\delta^{26}$Mg of CE-5 basalts are in the range of low-Ti lunar basalt (-0.263 ± 0.115‰) and are heavier than high-Ti basalts (-0.474 ± 0.198‰) (Sedaghatpour & Jacobsen 2019; Sedaghatpour et al. 2013).

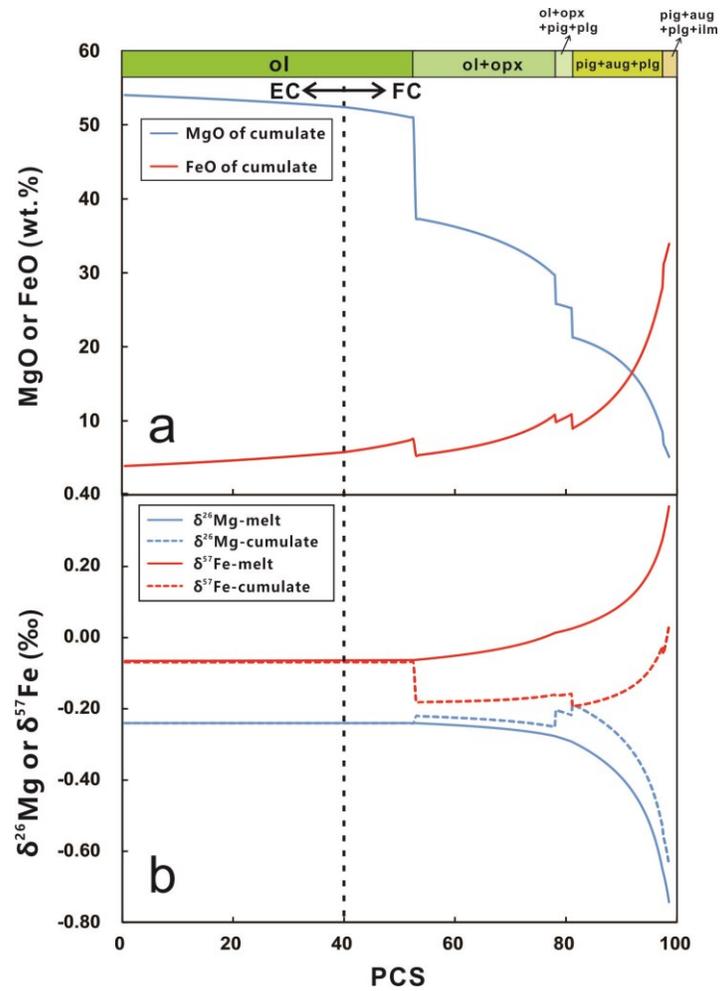

**Figure 3. Predicted FeO, MgO, δ⁵⁷Fe, and δ²⁶Mg evolution during LMO solidification.** The model is based on the magma ocean crystallization model in Charlier et al. (2018) with equilibrium crystallization (EC) up to 40% PCS of the LMO followed by fractional crystallization (FC). ol, olivine; opx, orthopyroxene; pig, pigeonite; aug, augite; plg, plagioclase; ilm, ilmenite. a. The FeO and MgO contents of cumulate during LMO differentiation; b. The solid line and dashed lines show evolution of Fe and Mg isotope compositions of residual melts and instantaneous cumulates, respectively.

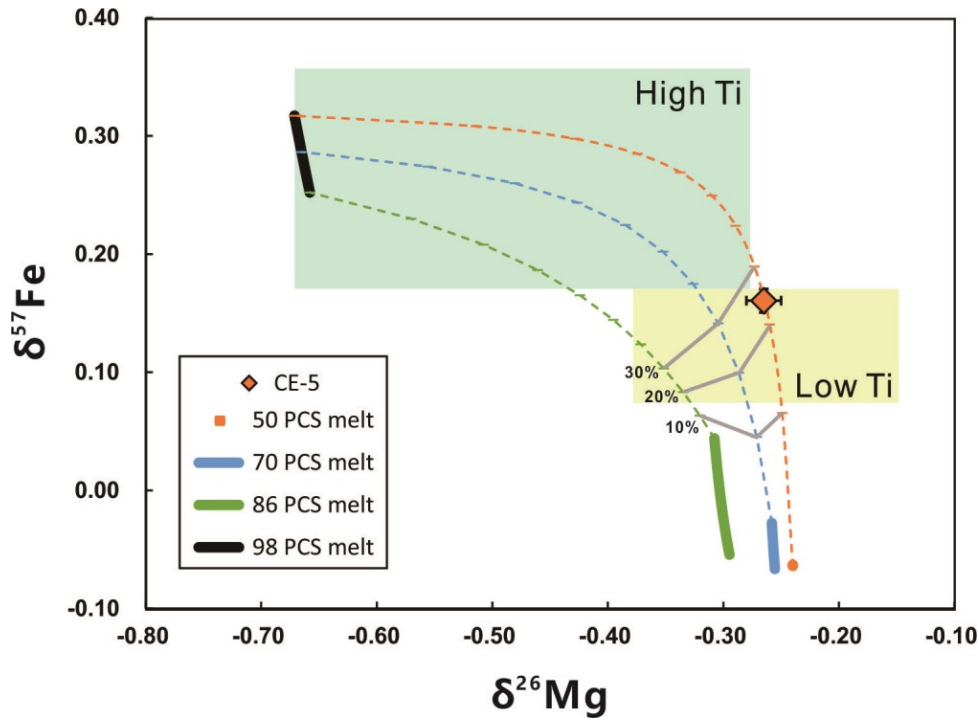

**Figure 4. Fe-Mg isotope mixing calculation of partial melts derived from 50, 70, 86, and 98 PCS at 1-20% melting degree.** The compositions for melts derived from 50 PCS (orange point), 70 PCS (blue bar), 86 PCS (green bar) and 98 PCS cumulates (black bar) at melting degree of 1-20% are calculated by a congruent batch melting model. The orange dashed line represents the mixing curve between 1% partial melts of 98 PCS cumulate and 1% partial melts of 50 PCS cumulate; The blue dashed line represents the mixing curve between 10% partial melts of 98 PCS cumulate and 1% partial melts of 70 PCS cumulate; The green dashed line represents the mixing curve between 20% partial melts of 98 PCS cumulate and 1% partial melts of 86 PCS cumulate. The gray solid lines represent mixing ratios of 98 PCS cumulate as 10%, 20%, and 30%. The green area represents the range of high-Ti basalt ($\delta^{57}$Fe: 0.264 ± 0.093‰ and $\delta^{26}$Mg: -0.474 ± 0.198‰) and the yellow area represents the range of low-Ti basalt ($\delta^{57}$Fe: 0.125 ± 0.051‰ and $\delta^{26}$Mg: -0.263 ± 0.115‰). Mg isotope data source (Sedaghatpour & Jacobsen 2019; Sedaghatpour et al. 2013), Fe isotope data source (Liu et al. 2010; Poitrasson et al. 2004; Sossi & Moynier 2017; Weyer et al. 2005).

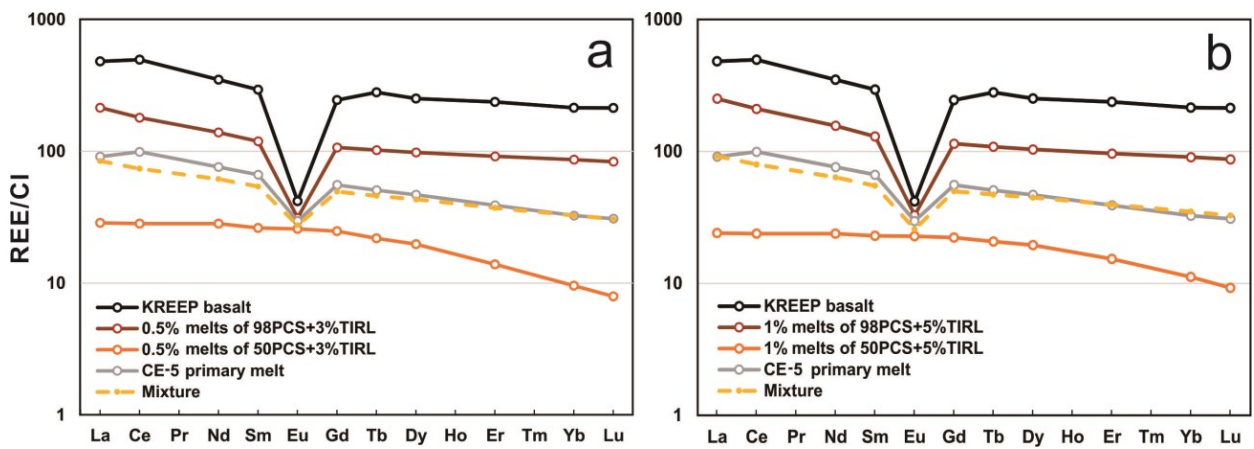

**Figure 5. REE mixing modeling between early cumulate melt and late cumulate melt.** The mixing between early cumulate melt and late cumulate melt can reproduce the REE contents of CE-5 basalt by selecting appropriate parameters, such as: (a) melts of 0.5% partial melting of an early cumulate (50 PCS + 3% TIRL) mixed with melts of 0.5% partial melting of a late cumulate (98 PCS + 3% TIRL) by a ratio of 70: 30 (mass ratio); (b) melts of 1% partial melting of an early cumulate (50 PCS + 5% TIRL) mixed with melts of 1% partial melting of a late cumulate (98 PCS + 5% TIRL) by a ratio of 70: 30. The REE contents of KREEP basalt are the value of SAU 169 (Gnos et al. 2004). CI chondrite compositions are from Barrat et al. (2012). The REE compositions of CE-5 primary magma are calculated by assuming that it underwent 38% fractional crystallization of olivine.

We thank the China National Space Administration for providing the CE-5 lunar samples used in this study. We thank Jiawei Li and Zaicong Wang for the help of ICP-MS analyses. This work was supported by pre-research Project on Civil Aerospace Technologies No. D020202 and D020302 funded by Chinese National Space Administration, the Strategic Priority Research Program of Chinese Academy of Sciences (XDB 41000000), the National Key Research and Development Program of China (2021YFA0716100), the National Natural Science Foundation of China (42173044, 42241146, 42073060, 41973060), the Key Research Program of the Chinese Academy of Sciences (ZDBS-SSW-JSC007), and the Minor Planet Foundation of China.

## Appendix

### A.1. Petrography and mineral chemistry

Using HR-XRTM via a Zeiss Xradia 520 Versa at the Nanjing Institute of Geology and Palaeontology, Chinese Academy of Sciences (NIGPAS), we first examined two basalt clasts (CE-5-01: 4.70 mg; CE-5-02: 6.50 mg). Due to the linear relationship between mineral density and its X-ray attenuation, it is easy to distinguish minerals of different morphology and density at micron-scale resolution tomographic slices. High-resolution 3D tomography observations show that two clasts were unaffected by the impact process.

Under a binocular microscope, representative slivers were then picked up, embedded in epoxy mounts, polished, and carbon coated for petrography and mineral chemistry. The petrological observations were carried out on a Hitachi S-3400 N scanning electron microscope equipped with an Oxford INCA 7021 energy dispersive spectroscope at Purple Mountain Observatory (PMO), CAS. The major element concentrations of mineral phases were determined via a JEOL JXA-8230 electron microprobe at PMO, CAS, using a beam current of 20 nA, an accelerating voltage of 15 kV and a focused electron beam. Both synthetic and natural mineral standards were used for calibration, and matrix corrections were based on ZAF procedures.

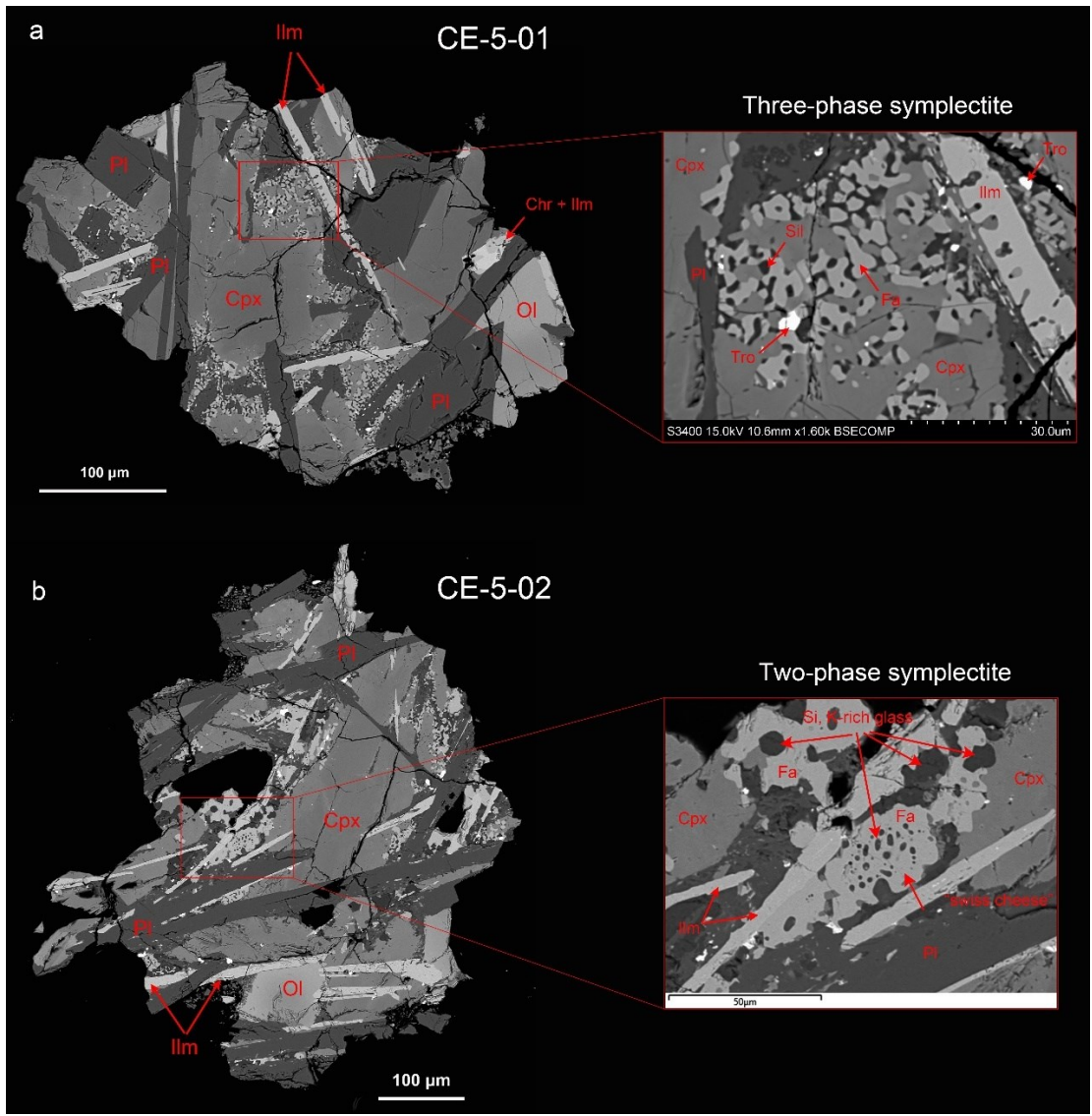

**Figure A1. Back scattered electron (BSE) images illustrating textures and mineralogy of CE-5-01 and CE-5-02.** Both two clasts show sub-ophitic textures and are dominated by clinopyroxene (Cpx) and plagioclase (Pl), with minor amounts of ilmenite (Ilm), olivine (Ol), chromite (Chr), and troilite (Tro). Symplectites consisting of two- and three-phase assemblages are abundant. **a.** The coexisting fayalite (Fa), hedenbergitic pyroxene and silica phase (Sil) suggest that they formed from the breakdown of pyroxferroite upon cooling near the lunar surface (Oba & Kobayashi 2001). **b.** The coexisting fayalite (Fa) and Si, K-glass illustrating a "swiss cheese" texture has generally been used as an indicator of highly fractionated and evolved basalts (Day et al. 2006).

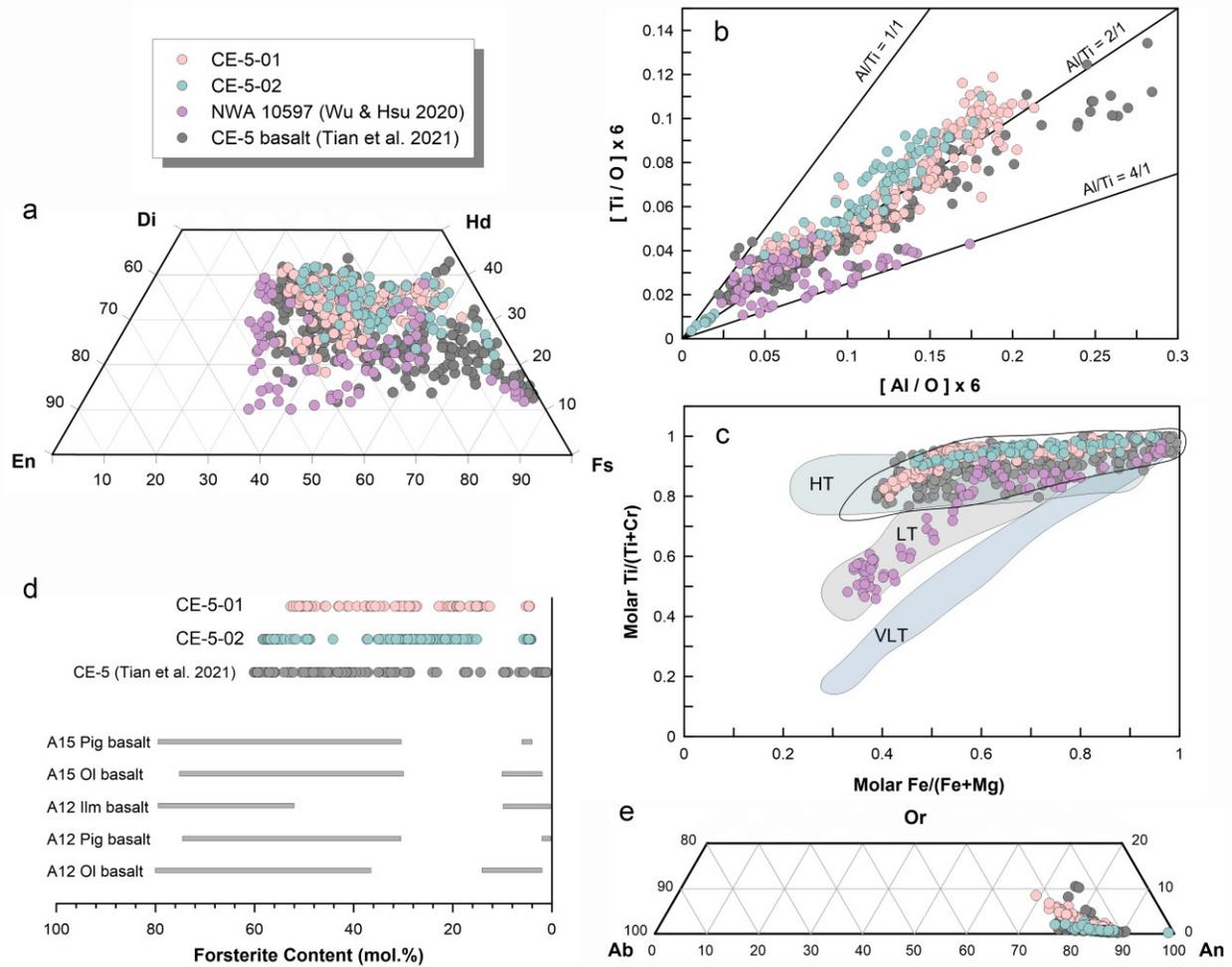

**Figure A2. Mineral chemistry of pyroxene (a–c), olivine (d), and plagioclase (e) in CE-5-01 and CE-5-02.** Low-Ti basalt meteorite NWA 10597 is shown for comparison (Wu & Hsu 2020). Two basalt clasts in this work (pink and green solids) are quite consistent in mineral chemistry with ones (gray solid) reported by Tian et al. (2021). Pyroxene displays a fractional crystallization trend distinct from that in NWA 10597 but shows an intermediate trend between low-titanium (LT) and high-titanium (HT) indicated by Apollo and Luna mare basalts (Robinson, Treiman, & Joy 2012). Di, diopside; Hd, hedenbergite; En, enstatite; Fs, ferrosilite. VLT, very-low-titanium; Or, orthoclase; Ab, albite; An, anorthite.

### A.2. Bulk major and trace elements analysis

Remaining basaltic fragments (CE-5-01: 4.39 mg; CE-5-02: 6.00 mg) were dissolved for bulk-rock major and trace elements analyses, with an Agilent 7700x ICP-MS at the State Key Laboratory of Geological Processes and Mineral Resources, China University of Geosciences, Wuhan. Due to the preciousness and limitation of the CE-5 samples, the analytical method of high accuracy and minimum consumption of ~ 5 mg samples was adopted. The detailed sample digestion, measurement procedures, operating conditions, and acquisition parameters of ICP-MS have been described by Zong et al. (2022). For monitoring the data quality, a low-Ti lunar meteorite NWA 10597 (4.90 mg), a primitive carbonaceous chondrite (Allende: 4.95 mg), a basaltic reference material (BCR-2: 5.39 mg) and a procedural blank were prepared and analyzed along with the CE-5 basalts in the same batch. A total of 46 major and trace element concentrations were reported in Table A1. Relative errors were lower than 5% and 10% for most major and trace elements, respectively. About 1 mg sample was consumed during the ICP-MS analysis.

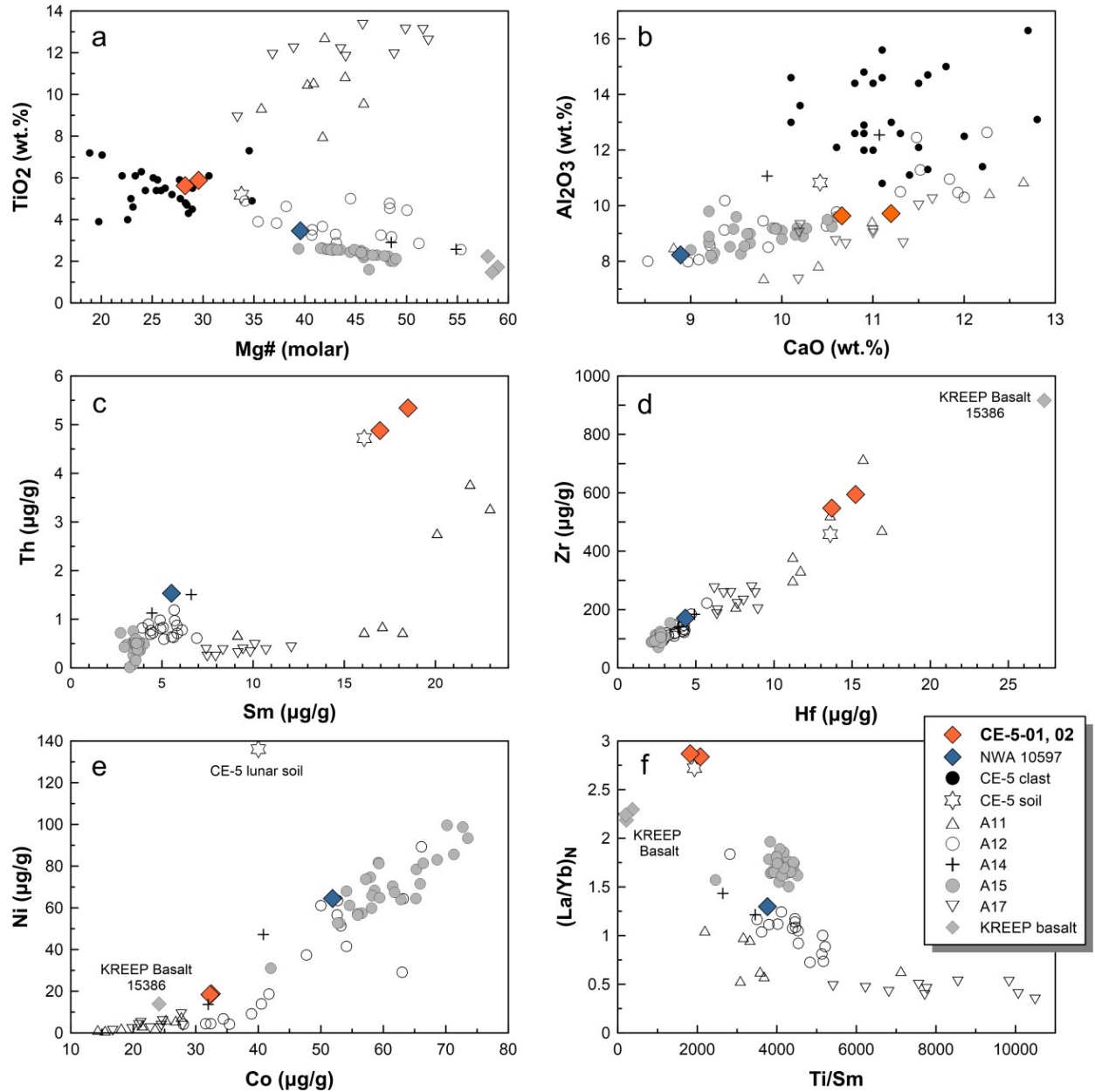

**Figure A3. Bulk major and trace element compositions of CE-5-01, CE-5-02.** CE-5 clast (Su et al. 2022), CE-5 soil (Yao et al. 2022), A11, 12, 14, 15, 17 basalts (Elardo et al. 2014), as well as three KREEP basalt meteorites (Gnos et al. 2004; Neal & Kramer 2003; Warren 1989) are shown for comparison. CI chondrite composition is from Barrat et al. (2012). CE-5 basalt has $TiO_2$ contents lower than A11 and A17 high-Ti basalts (8.0–13.3 wt.%), but slightly higher than A12 and A15 low-Ti basalts (1.6–4.5 wt.%). Low Mg#, moderate $Al_2O_3$ and CaO, low abundances of compatible elements (e. g., Ni, Co) make them fall within the fields defined by A12 and A15 basalts. Incompatible

trace element concentrations (e.g., Th, Sm, Zr, Hf) are significantly higher than those in A12 and A15 basalts, but lower than those in KREEP basalts, such as Th (15.4 μg/g) and Sm (36.5 μg/g) for KREEP basalt 15386 (Neal & Kramer 2003).

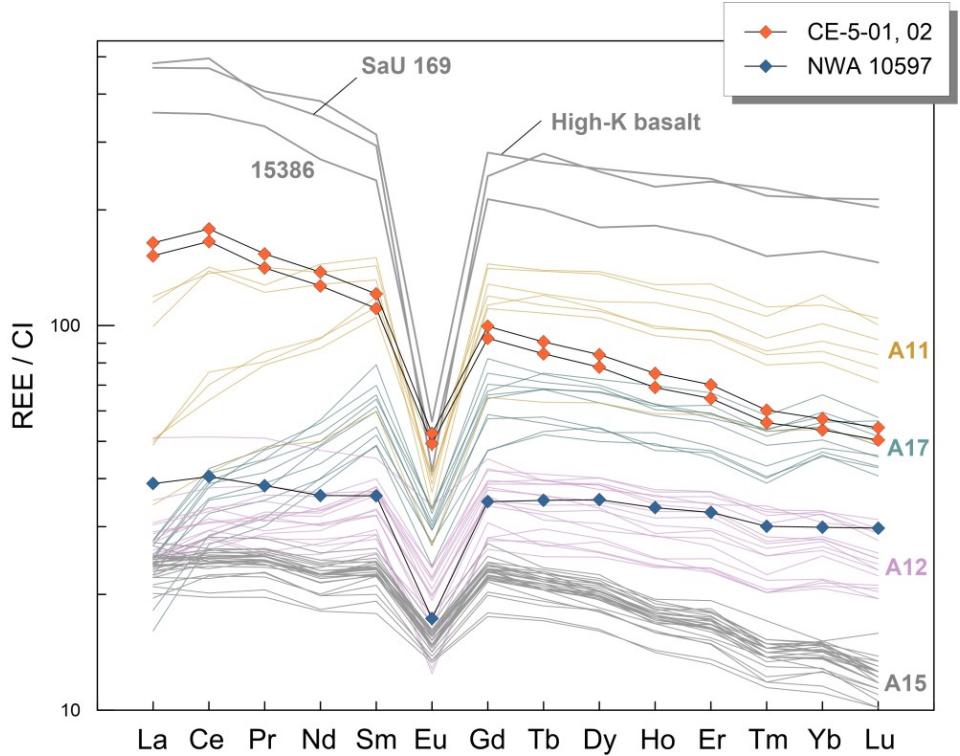

**Figure A4. Chondrite-normalized REE plots for CE-5-01, CE-5-02.** A11, 12, 15, 17 basalts (Clive Neal's Mare Basalt database, https://www3.nd.edu/~cneal/Lunar-L/) and three KREEP basalts (15386, high-K basalt and SaU 169) (Gnos et al. 2004; Neal & Kramer 2003; Warren 1989) are also shown for comparison. CI chondrite composition is from Barrat et al. (2012). CE-5 basalt displays a KREEP-like REE pattern, with a high LREE/HREE, a high (La/Sm)$_N$ of 1.37 (average value), a high (La/Yb)$_N$ of 2.86, and a deep negative Eu anomaly (0.49).

**Table A1**. Major and trace elements analyses of CE-5 basalts and standards by ICP-MS.

| Sample Name | CE-5-01 | CE-5-02 | NWA 10597 | Allende | BCR-2 | BCR-2 | Blank |
|---|---|---|---|---|---|---|---|
|  | 4.39 mg | 6.00 mg | 4.90 mg | 4.95 mg | 5.39 mg | Reference value* |  |
| **Major elements** | **wt.%** | **wt.%** | **wt.%** | **wt.%** | **wt.%** | **wt.%** | **ng/g** |
| SiO$_2$ | 41.68 | 42.11 | 41.70 |  | 55.84 | 54.1 |  |
| TiO$_2$ | 5.88 | 5.62 | 3.47 | 0.12 | 2.25 | 2.26 | 1.59 |
| Al$_2$O$_3$ | 9.72 | 9.63 | 8.23 | 2.71 | 13.41 | 13.5 | 1.02 |
| Cr$_2$O$_3$ | 0.21 | 0.20 | 0.38 | 0.55 | 0.002 | 0.002 | 0.14 |
| FeO | 24.32 | 24.99 | 26.68 | 31.59 | 12.24 | 12.42 | 30.8 |
| MnO | 0.30 | 0.29 | 0.32 | 0.18 | 0.20 | 0.20 | 0.12 |
| MgO | 5.73 | 5.52 | 9.80 | 24.43 | 3.59 | 3.59 | 0.32 |
| CaO | 11.20 | 10.66 | 8.89 | 2.34 | 7.21 | 7.12 | 5.75 |
| Na$_2$O | 0.47 | 0.47 | 0.31 | 0.41 | 3.11 | 3.16 | 12.6 |
| K$_2$O | 0.21 | 0.22 | 0.13 | 0.05 | 1.79 | 1.79 | 73.7 |
| P$_2$O$_5$ | 0.28 | 0.29 | 0.09 | 0.24 | 0.35 | 0.35 | 14.7 |
| Mg# | 29.6 | 28.2 | 39.6 | | | | |
| **Trace elements** | µg/g | µg/g | µg/g | µg/g | µg/g | µg/g | ng/g |
| Li | 14.6 | 14.9 | 10.5 | 1.66 | 9.04 | 9.00 | 0.15 |
| Be | 2.68 | 2.78 | 0.92 | 0.037 | 1.94 | 2.17 | 0.000 |
| Sc | 65.2 | 60.5 | 49.1 | 9.48 | 33.0 | 33.0 | 0.013 |
| V | 90.0 | 87.7 | 102 | 80.5 | 419 | 416 | 0.56 |
| Cr | 1435 | 1375 | 2579 | 3771 | 16.3 | 16.5 | 0.14 |
| Co | 32.5 | 32.3 | 51.9 | 724 | 36.7 | 37.0 | 0.0054 |
| Ni | 18.8 | 18.4 | 64.4 | 15094 | 13.3 | 13.0 | 0.023 |
| Cu | 14.7 | 15.8 | 10.8 | 105 | 18.1 | 18.4 | 0.025 |
| Zn | 10.2 | 9.13 | 7.12 | 97.5 | 129 | 133 | 0.14 |
| Ga | 5.88 | 6.06 | 3.73 | 5.73 | 22.6 | 23.0 | 0.0006 |
| Rb | 4.77 | 5.03 | 2.52 | 1.14 | 46.4 | 46.9 | 0.0047 |
| Sr | 342 | 360 | 120 | 12.7 | 340 | 340 | 0.0042 |
| Y | 112 | 118 | 51.3 | 2.46 | 36.3 | 37.0 | 0.0003 |
| Zr | 548 | 594 | 170 | 6.01 | 196 | 184 | 0.0073 |
| Nb | 35.2 | 37.3 | 12.4 | 0.49 | 12.4 | 12.6 | 0.0035 |
| Cs | 0.20 | 0.22 | 0.056 | 0.078 | 1.10 | 1.10 | 0.0005 |
| Ba | 394 | 424 | 132 | 4.95 | 694 | 677 | 0.0061 |
| La | 35.7 | 38.6 | 9.12 | 0.45 | 24.5 | 24.9 | 0.0005 |
| Ce | 99.2 | 107 | 24.3 | 1.12 | 51.7 | 52.9 | 0.0014 |
| Pr | 12.8 | 14.0 | 3.49 | 0.17 | 6.76 | 6.70 | 0.0002 |
| Nd | 58.8 | 63.9 | 16.8 | 0.82 | 29.0 | 28.7 | 0.0000 |
| Sm | 17.0 | 18.5 | 5.52 | 0.26 | 6.76 | 6.58 | 0.0000 |
| Eu | 2.89 | 3.07 | 1.01 | 0.097 | 1.90 | 1.96 | 0.0000 |
| Gd | 19.1 | 20.5 | 7.17 | 0.34 | 6.66 | 6.75 | 0.0003 |
| Tb | 3.17 | 3.40 | 1.32 | 0.064 | 1.04 | 1.07 | 0.0000 |
| Dy | 19.8 | 21.3 | 8.96 | 0.43 | 6.60 | 6.41 | 0.0000 |
| Ho | 3.91 | 4.25 | 1.90 | 0.096 | 1.30 | 1.28 | 0.0001 |
| Er | 10.7 | 11.6 | 5.42 | 0.29 | 3.63 | 3.66 | 0.0002 |
| Tm | 1.47 | 1.58 | 0.79 | 0.048 | 0.52 | 0.54 | 0.0001 |
| Yb | 9.00 | 9.61 | 5.02 | 0.30 | 3.29 | 3.38 | 0.0000 |
| Lu | 1.24 | 1.34 | 0.73 | 0.044 | 0.48 | 0.50 | 0.0000 |
| Hf | 13.7 | 15.2 | 4.33 | 0.16 | 4.83 | 4.90 | 0.0002 |
| Ta | 1.79 | 1.89 | 0.64 | 0.024 | 0.77 | 0.78 | 0.0004 |
| Pb | 1.49 | 1.67 | 1.05 | 1.14 | 9.87 | 10.85 | 0.023 |
| Th | 4.88 | 5.34 | 1.54 | 0.051 | 5.95 | 5.70 | 0.0005 |
| U | 1.25 | 1.36 | 0.40 | 0.016 | 1.64 | 1.69 | 0.0003 |

**Note.** Due to the loss of Si during the sample digestion procedure, SiO$_2$ concentration was obtained by subtraction of other major elements from 100 wt.%.

* Reference values of BCR-2 are from the GEOREM database (http://georem.mpch-mainz.gwdg.de/).

### A.3. Bulk Sr, Mg, and Fe isotopes analysis

The remaining solutions were analyzed for radiogenic Sr isotope, stable Mg and Fe isotope compositions at the CAS Key Laboratory of Crust-Mantle Materials and Environments at the University of Science and Technology of China (USTC) using a Neptune Plus MC-ICP-MS (Tables A2 and A3). First, the Sr aliquots were collected following the method developed by Chen et al. (2022). Briefly, Sr was purified using the AG50W-X12 resin (200–400 mesh). The matrix elements were collected for subsequent Fe–Mg isotope column chemistry by eluting with 2.5N HCl. Strontium was then collected with 4N HCl. To ensure Sr was fully separated from HREE and Hf, a second column was applied by using a polypropylene spin column containing 0.75 mL of AG50W-X12 resin. The total procedure blanks were lower than 100 pg, with negligible compared with the amount of Sr (1.2–3 μg) in the loaded samples. The instrument bias for $^{87}Sr/^{86}Sr$ was calibrated with $^{86}Sr/^{88}Sr = 0.1194$ (Nier 1938) by an exponential law. The $^{87}Sr/^{86}Sr$ obtained for BCR-2 is $0.705012 \pm 0.000030$, consistent with the literature value (e.g., $0.705015 \pm 0.000013$) (Balcaen et al. 2005).

The Mg aliquots were obtained by passing the matrix solutions collected in the Sr procedure using AG50W-X12 resin (200–400 mesh). The chemical purification procedure for Mg was described by An et al. (2014). To obtain a pure Mg solution, the same procedure was conducted twice. The matrix elements were collected for subsequent Fe isotope column chemistry by eluting with $2N\ HNO_3 + 0.5N\ HF$. Magnesium was then collected with $1N\ HNO_3$. The yields in the column chemistry were generally above 99%. Procedural blanks for Mg were less than 10 ng, which is insignificant relative to the amount of Mg loaded on the column ($\geq 40$ μg). Measurements were done using the sample-standard-bracketing method. Magnesium isotope compositions were reported as: $\delta^{26}Mg = [(^{26}Mg/^{24}Mg)_{sample}/(^{26}Mg/^{24}Mg)_{DSM-3} -1]$ (‰). The $\delta^{26}Mg$ obtained for BCR-2 and Allende are $-0.189 \pm 0.027$‰ (2SD, N=6) and $-0.274 \pm 0.015$‰ (2SD, N=9), respectively, which are comparable to literature data within analytical error (e.g., $-0.25 \pm 0.06$‰ for BCR-2 in Pogge von Strandmann et al. (2011); $-0.30 \pm 0.05$‰ for Allende in Teng et al. (2010)).

Iron was purified using Bio-Rad AG1-X8 resin following the method described in An et al. (2017). Briefly, matrix elements were removed by washing with 6N HCl (washes were collected to assess the loss of Fe for yield check). Iron was eluted using 0.4N HCl and $H_2O$ followed by 6N HCl with yields greater than 99%. Procedural blanks for Fe were less than 10 ng, which is insignificant relative to the amount of Fe put through chemical purification ($\geq 60$ μg). Iron isotope ratios were reported relative to IRMM-014 in δ notation: $\delta^{57}Fe = [(^{57}Fe/^{54}Fe)_{sample}/(^{57}Fe/^{54}Fe)_{IRMM-014} - 1]$ (‰). The $\delta^{57}Fe$ obtained for BCR-2 and Allende are $0.099 \pm 0.058$‰ (2SD, N=3) and $-0.030 \pm 0.041$‰ (2SD, N=14), respectively, which are comparable to literature data within analytical error (e.g., $0.12 \pm 0.03$‰ for BCR-2 in He et al. (2015); $0.003 \pm 0.019$‰ for Allende in Craddock & Dauphas (2011)).

**Table A2**. Rb-Sr isotope data of CE-5 basalts and standards.

| Sample | CE-5-01 | CE-5-02 | BCR-2 |
| --- | --- | --- | --- |
| $^{87}Rb/^{86}Sr$ | 0.03942 | 0.03942 | |
| $^{87}Sr/^{86}Sr$ | 0.700916 | 0.700918 | 0.705012 |
| 2SD | 0.00003 | 0.00003 | 0.000025 |
| $^{87}Sr/^{86}Sr$ (i) | 0.69976 | 0.69977 | |
| 2SD | 0.00003 | 0.00003 | |
| $^{87}Rb/^{86}Sr$ of source region | 0.02005 | 0.0201 | |

The initial $^{87}Sr/^{86}Sr$ ratios (i.e., $^{87}Sr/^{86}Sr$ (i)) of CE-5 basalts are calculated based on the Pb-Pb age of 2.03 Ga (Li et al. 2021), which are consistent with results obtained by Tian et al. (2021) (0.69915 to 0.69986). The $^{87}Rb/^{86}Sr$ ratios of CE-5 basalts source region are calculated by assuming a single-stage evolution model with differentiated age of 4.56 billion years and an initial $^{87}Sr/^{86}Sr$ = 0.69903 and $^{87}Rb/^{86}Sr$ = 2.71828 (Nyquist 1977).

**Table A3.** Mg and Fe isotopes data (‰) of CE-5 basalts and standards.

| Sample | CE-5-01 | CE-5-02 | Allende | BCR-2 | IGG | USTC-Fe |
|---|---|---|---|---|---|---|
| Description | Lunar basalt | Lunar basalt | CV3 chondrite | USGS standard | Internal standard | Internal standard |
| $\delta^{25}Mg$ | -0.129 | -0.145 | -0.14 | -0.099 | -0.909 | |
| 2SD | 0.024 | 0.03 | 0.032 | 0.01 | 0.017 | |
| 2SE | 0.006 | 0.008 | 0.011 | 0.004 | 0.006 | |
| $\delta^{26}Mg$ | -0.246 | -0.282 | -0.274 | -0.189 | -1.75 | |
| 2SD | 0.024 | 0.027 | 0.015 | 0.027 | 0.028 | |
| 2SE | 0.006 | 0.007 | 0.005 | 0.011 | 0.009 | |
| N | 15 | 15 | 9 | 6 | 9 | |
| $\delta^{56}Fe$ | 0.114 | 0.099 | -0.033 | 0.061 | | 0.695 |
| 2SD | 0.027 | 0.023 | 0.026 | 0.027 | | 0.038 |
| 2SE | 0.007 | 0.006 | 0.007 | 0.016 | | 0.013 |
| $\delta^{57}Fe$ | 0.172 | 0.149 | -0.03 | 0.099 | | 1.029 |
| 2SD | 0.049 | 0.042 | 0.041 | 0.058 | | 0.06 |
| 2SE | 0.013 | 0.011 | 0.011 | 0.033 | | 0.02 |
| N | 15 | 15 | 14 | 3 | | 9 |

N denoted the number of repeated analyses.

The long-term average $\delta^{26}Mg$ of IGG is -1.743 ± 0.044‰ (2SD, N=2764). The long-term average $\delta^{56}Fe$ of USTC-Fe is 0.700 ± 0.030‰ (2SD, N=100).

### A.4. LMO differentiation model

The differentiation history of the lunar magma ocean (LMO) has been extensively studied by experiments and thermodynamic modeling (Charlier et al. 2018; Elardo, Draper, & Shearer 2011; Elkins-Tanton, Burgess, & Yin 2011; Johnson et al. 2021; Lin et al. 2017; Rapp & Draper 2018). Our LMO modeling is inspired by Klaver et al. (2021), who used the experimental results of Charlier et al. (2018) to predict the Ca isotope evolution during LMO solidification. In this study, the model of Charlier et al. (2018) was selected as the basis for the modeling of major elements, REE contents, and Fe-Mg isotope evolution. This selection is based on two reasons, which have been explained by Klaver et al. (2021): 1) the experiments in Charlier et al. (2018) cover the entire solidification interval for a LPUM bulk Moon composition (Longhi 2006); and 2) Charlier et al. (2018) present equations for the crystallized mineral composition as a function of pressure, temperature and melt chemistry, thus simplify the forward modeling of LMO solidification. Previous Fe-Mg-Ca isotope studies for lunar basalt have emphasized that the selection of other LMO models will not show a significantly different isotope evolution (Klaver et al. 2021; Sedaghatpour & Jacobsen 2019). The crystalized plagioclase is assumed to float and form the lunar anorthosite crust and the minerals with a density greater than the melt (olivine, pyroxene, and ilmenite) are presumed to sink and form the lunar cumulate layer. According to previous studies of Snyder, Taylor, & Neal (1992) and Van Orman & Grove (2000), the sinking instantaneous cumulates contain a small fraction (7 wt.%) of plagioclase that is trapped between the mafic phases. The evolution of the major elements of the melt and crystallized mineral was calculated following the equation of phase compositions of Charlier et al. (2018) in 0.5 wt.% (up to 60% LMO solidification) and 0.2 wt.% (60–99% LMO solidification) crystallization increments. Cotectic phase proportions were also taken from Charlier et al. (2018).

The Fe-Mg isotope compositions of each increment was calculated based on the phase proportions and FeO-MgO of cumulate and residual melt. The calculation for δ value of melt during equilibrium crystallization follows:

$$\delta_{melt} = \frac{1000 + \delta_{total}}{f_{melt} + (1 - f_{melt}) \times a_{solid-melt}} - 1000$$

For fractional crystallization:

$$\delta_{melt} = [1000 + \delta_{total}] \times [f_{melt}^{(a_{solid-melt}-1)} - 1000]$$

where $\delta_{melt}$ refers to the isotope composition of residual melt, $\delta_{total}$ is the composition of the bulk system (i.e., the BSM value for equilibrium crystallization or the melt value of the last increment for fractional crystallization), $f_{melt}$ refers to the mass fraction of Fe or Mg in the melt relative to the bulk system, $a_{solid-melt}$ refers to the bulk equilibrium fractionation factor between bulk solid and melt, which can be calculated:

$$1000 ln a_{solid-melt} = \frac{\sum F_{mineral} \times C_{mineral} \times 1000 ln a_{mineral-melt}}{\sum F_{mineral} \times C_{mineral}}$$

$F_{mineral}$ refers to the proportion of mineral phases and $C_{mineral}$ is the Fe or Mg content of minerals. The isotope compositions of cumulate are calculated as:

$$\delta_{cumulate} = \delta_{melt} + 1000 ln a_{solid-melt}$$

The $1000 ln\alpha_{mineral-melt}$ is a function of crystallizing temperature. Feldspar contains negligible Fe and Mg. It does not affect the Fe-Mg isotope evolution of LMO and is not taken into consideration for isotope calculation. For olivine, pyroxene, and ilmenite, we used the Fe force constants of minerals to calculate equilibrium Fe isotope fractionation as $1000 ln\alpha_{mineral-melt} = 2904 \times (F_{mineral} - F_{melt})/T^2$, where F is the force constant and T is crystallization temperature in Kelvin. Force constants are from references (Dauphas et al. 2012; Dauphas et al. 2014; Nie et al. 2021) and are listed in Table A4:

**Table A4.** Iron force constants used for calculation of LMO differentiation.

| Phases | Fe force constant (N/m) |
|---|---|
| Basaltic melt | 199 |
| Olivine | 197 |
| Pyroxene | 150 |
| Ilmenite | 133 |

We used the reduced partitioning function ratios (β value) obtained by first-principles calculations to calculate the Mg isotope equilibrium fractionation between minerals and melts as: $1000\ln\alpha_{mineral-melt}=1000\ln\beta_{mineral}-1000\ln\beta_{melt}$. The $1000\ln\beta$ of minerals and melts can be calculated by: $1000\ln\beta = a \times 10^6/T^2 + b \times (10^6/T^2)^2 + c \times (10^6/T^2)^3$. T is the crystalizing temperature in Kelvin and a, b and c are polynomial fitting parameters reported by Wang et al. (2023) and references therein (Table A5):

**Table A5**. Calculation of Mg isotope reduced partitioning function ratios for minerals and melts.

| Phases | a | b | c |
| --- | --- | --- | --- |
| Basaltic melt | 2.565 | -0.01735 | 0.000195 |
| Olivine | 2.565 | -0.01735 | 0.000195 |
| Orthopyroxene | 2.641 | -0.01878 | 0.000214 |
| Clinopyroxene | 2.767 | -0.01812 | 0.000188 |
| Ilmenite | 2.334 | -0.0126 | 0.000107 |

The $\delta^{26}$Mg composition of the bulk silicate moon (BSM) used for calculation is -0.24±0.06‰, which is consistent with the composition of bulk silicate earth (BSE) (Sedaghatpour & Jacobsen 2019; Sedaghatpour et al. 2013). The $\delta^{57}$Fe of BSM adopts the value estimated by Elardo et al. (2019) of -0.065±0.055‰.

The REE evolution at each increment was calculated based on the phase proportions of the LMO modeling. The REE content of residual melt during equilibrium crystallization is:

$$C_{melt} = \frac{C_{total}}{f \times D + (1-f)}$$

And for fractional crystallization:

$$C_{melt} = C_{total} \times (1-f)^{D-1}$$

where $C_{total}$ represents the composition of the bulk system (i.e., the BSM composition for equilibrium crystallization or the residual melt composition of last increments for fractional crystallization), $f$ refers to the mass fraction of minerals crystalized from the melt and $D$ refers to bulk partition coefficients:

$$D = \sum F_{mineral} \times D_{mineral}$$

where $F_{mineral}$ refers to the proportion of mineral phases and $D_{mineral}$ is the partition coefficient (Table A6). The REE contents of BSM used the values estimated by Warren (2005).

**Table A6.** The REE partition coefficients used in the calculation.

|    | Ol[a]  | Opx[b] | Aug[b] | Pig[c] | Plg[d] |
|----|--------|--------|--------|--------|--------|
| La | 0.0001 | 0.0007 | 0.0446 | 0.0009 | 0.0418 |
| Ce | 0.0001 | 0.0015 | 0.0733 | 0.0017 | 0.0302 |
| Nd | 0.0001 | 0.0055 | 0.1544 | 0.0058 | 0.0236 |
| Sm | 0.0006 | 0.0143 | 0.251  | 0.011  | 0.017  |
| Eu | 0.0007 | 0.0204 | 0.2952 | 0.0068 | 1.2    |
| Gd | 0.001  | 0.0281 | 0.3377 | 0.021  | 0.0105 |
| Tb | 0.002  | 0.0376 | 0.3758 | 0.027  | 0.0095 |
| Dy | 0.003  | 0.0487 | 0.4071 | 0.034  | 0.0089 |
| Er | 0.008  | 0.0714 | 0.4402 | 0.055  | 0.0077 |
| Yb | 0.019  | 0.0913 | 0.4426 | 0.087  | 0.0065 |
| Lu | 0.03   | 0.0995 | 0.4368 | 0.11   | 0.0068 |

[a] from McKay (1986). [b] from Yao, Sun, & Liang (2012). [c] from McKay, Le, & Wagstaff (1991). [d] from Phinney & Morrison (1990). [e] from Hauri, Wagner, & Grove (1994).